\documentclass[a4paper,12pt]{article}
\usepackage[utf8]{inputenc}
\usepackage{graphicx}
\usepackage{hyperref}
\usepackage{latexsym}
\usepackage{color}
\usepackage{amsmath}
\usepackage{geometry}
\usepackage{slashed}

\def\Tr{{\rm Tr}}
\def\msbar{{\overline{\rm MS}}}

\title{Quark chiral condensate from the overlap quark propagator}
\author{Chao Wang$^1$\thanks{wangchao88@ihep.ac.cn}, Yujiang Bi$^2$, Hao Cai$^2$, Ying Chen$^1$, Ming Gong$^1$,\\
Zhaofeng Liu$^1$\thanks{liuzf@ihep.ac.cn}}

\date{}

\begin{document}
\maketitle

\begin{center}
$^1$Institute of High Energy Physics and Theoretical Physics Center for Science Facilities, Chinese Academy of Sciences, Beijing 100049, China\\
$^2$School of Physics and Technology, Wuhan University, Wuhan 430072, China
\end{center}

\begin{abstract}
From the overlap lattice quark propagator calculated in the Landau gauge,
we determine the quark chiral condensate by fitting operator product expansion
formulas to the lattice data. The quark propagators are computed on domain wall fermion configurations
generated by the RBC-UKQCD Collaborations with $N_f=2+1$ flavors. Three ensembles with different
light sea quark masses are used at one lattice spacing $1/a=1.75(4)$ GeV.
We obtain $\langle\bar\psi\psi\rangle^\msbar(2\mbox{ GeV})=(-305(15)(21)\mbox{ MeV})^3$ in the SU(2) chiral limit.
\end{abstract}

\newpage
\section{Introduction}
The strong interactions among quarks and gluons have two prominent features
at low energies: confinement and chiral symmetry breaking.
The quark chiral condensate $\langle\bar\psi\psi\rangle$, which is in the light quark massless limit,
is the order parameter of the spontaneous chiral symmetry breaking in Quantum Chromodynamics (QCD),
the theory describing strong interaction.
Furthermore $\Sigma\equiv-\langle\bar\psi\psi\rangle$ is one of the two low energy constants
of chiral perturbation theory, the low energy effective theory of QCD, at leading order. 
The quark chiral condensate also appears in QCD sum rules and 
is an important input parameter.

Thus there have been many determinations of the chiral condensate from different ways 
by using lattice QCD, which is the nonperturbative method to solve QCD from first principles. 
See, for examples, 
Refs~\cite{Blum:2014tka,Bazavov:2010yq,Cichy:2013gja,Borsanyi:2012zv,Durr:2013goa,Baron:2009wt,Brandt:2013dua,Engel:2014eea,DeGrand:2006nv}.
A review of the evaluations of the chiral condensate on the lattice can be found in Ref.~\cite{Aoki:2016frl}.

In this work, we determine the SU(2) low energy constant $\Sigma$ by comparing the Operator Product Expansion (OPE) of the quark
propagator in momentum space in the continuum $\msbar$ scheme with the lattice calculation of the propagator in Landau gauge. This strategy was
used by the ETM Collaboration in a calculation with two flavors of dynamical Wilson twisted mass fermions~\cite{Burger:2012ti}.
Our analysis is based on 2+1-flavor domain wall
fermion configurations and overlap valence quarks.
There were also analysis using the staggered fermions~\cite{Bowman:2006zk}, the OPE of the pseudoscalar vertex~\cite{Becirevic:2004qv,Boucaud:2009kv}
and the OPE of the quark propagator in coordinate space~\cite{Gimenez:2005nt}.

Our final result 
obtained at one lattice spacing is $\Sigma^{1/3}=305(15)(21)$ MeV
in the $\msbar$ scheme at the renormalization scale 2 GeV.
Here the first error contains uncertainties from statistics, the lattice spacing and truncation effects in perturbative calculations.
The second error is an estimation of the $\mathcal{O}(a^2g^2)$ lattice artifacts in our data.

In the rest of the paper, we first discuss the OPE of the quark propagator in the $\msbar$ scheme in Sec.~\ref{sec:ope_prop}.
Then our lattice setup is given in Sec.~\ref{sec:lattice}. The analysis of the quark propagator and the results
of the chiral condensate are presented in Sec.~\ref{sec:analysis}. Finally we summarize in Sec.~\ref{sec:summary}.

\section{OPE of the quark propagator}
\label{sec:ope_prop}
For a quark field $\psi$ with mass $m_q$, its propagator in momentum space $S_q(p)$ can be written as
\begin{equation}
S_q(p)\equiv\int dx e^{-ipx}\langle T\psi(x)\bar\psi(0)\rangle=\frac{-i\slashed p V(p^2)}{p^2}+\frac{S(p^2)}{p^2},
\label{eq:qprop}
\end{equation}
where the dressing functions $S(p^2)$ and $V(p^2)$ will be called the scalar and vector form factor respectively at below.
The OPE of these two form factors renormalized in the $\msbar$ scheme in Landau gauge was calculated to three loops in Ref.~\cite{Chetyrkin:2009kh}.
Up to operators of dimension three, one has
\begin{eqnarray}
S_R(p^2) &=& S_{PT}(\mu,p^2)m_q(\mu) + \frac{C_{m^3}(\mu,p^2)}{p^2}m_q^3 + \frac{C_{mA^2}(\mu,p^2)}{p^2}\langle m_q A^2\rangle  \nonumber\\
&&+ \frac{C_{\bar\psi\psi}(\mu,p^2)}{p^2}\langle\bar\psi\psi\rangle(\mu),
\label{eq:SR}
\end{eqnarray}
and
\begin{equation}
V_R(p^2)=V_{PT}+\frac{C_{m^2}(\mu,p^2)}{p^2}m_q^2+\frac{C_{A^2}(\mu,p^2)}{p^2}\langle A^2\rangle.
\label{eq:VR}
\end{equation}
Here the purely perturbative parts $S_{PT}$ and $V_{PT}$ were computed at three loops in Ref.~\cite{Chetyrkin:1999pq}. The Wilson coefficients $C_{m^3}$,
$C_{mA^2}$, $C_{\bar\psi\psi}$, $C_{m^2}$ and $C_{A^2}$ at three loops can be found in Ref.~\cite{Chetyrkin:2009kh}.

In principle if we can obtain the scalar and vector form factors by lattice QCD,
then we can fit the lattice data to the functions in Eqs.(\ref{eq:SR},\ref{eq:VR}) to extract out
the quark mass and the chiral condensate.
Since we need the inverse powers of $p^2$ to suppress the contributions from higher dimension operators,
the lower limit of the fitting range in $p^2$ can not be too small. The Wilson coefficients are calculated by perturbation theory. This also
requires $p^2$ can not be too small. On the other hand, if $p^2$ is too large then $\mathcal{O}(a^2p^2)$ and higher order lattice discretization effects in
the data will be out of control.
Thus one needs to find a fitting window in which a stable and reliable value for the chiral condensate can be obtained.

Before the fittings we do not know if such a window exists or not given the lattice spacing in our data. 
Therefore we will vary our fitting range to test the reliability of our results.
And we shall take into account the lattice discretization artifacts in our error analysis.

\section{Lattice setup}
\label{sec:lattice}
We use the 2+1-flavor domain wall fermion configurations generated by the RBC-UKQCD
collaborations~\cite{Aoki:2010dy}. The parameters of the ensembles used
in this analysis are given in Tab.~\ref{tab:confs}.
\begin{table}
\begin{center}
\caption{Parameters of the 2+1-flavor domain wall fermion configurations generated by the RBC-UKQCD collaboration.
The residual mass is from Ref.~\cite{Aoki:2010dy}.
The lattice spacing was determined in Ref.~\cite{Yang:2014sea}.}
\begin{tabular}{cccccc}
\hline\hline
$1/a$(GeV) & label & $am^{sea}_l/am^{sea}_s$ & volume & $N_{conf}$ & $am_{res}$\\
\hline
1.75(4) & c005  & 0.005/0.04 & $24^3\times64$ & $92$ & 0.003152(43) \\
& c01 & 0.01/0.04 & $24^3\times64$ & $88$ & \\
& c02 & 0.02/0.04 & $24^3\times64$ & $138$ & \\
\hline\hline
\end{tabular}
\label{tab:confs}
\end{center}
\end{table}
Three light sea quark masses are used to check the sea quark
mass dependence of our results.

We use overlap fermions for the valence quark.
The massless overlap operator~\cite{Neuberger:1997fp} is defined as
\begin{equation}
D_{ov}  (\rho) =   1 + \gamma_5 \varepsilon (\gamma_5 D_{\rm w}(\rho)).
\end{equation}
Here $\varepsilon$ is the matrix sign function and $D_{\rm w}(\rho)$ is the usual Wilson fermion operator,
except with a negative mass parameter $- \rho = 1/2\kappa -4$ in which $\kappa_c < \kappa < 0.25$.
In our calculation we use $\kappa = 0.2$ which corresponds to $\rho = 1.5$. The massive overlap Dirac operator is defined as
\begin{eqnarray}
D_m &=& \rho D_{ov} (\rho) + m\, (1 - \frac{D_{ov} (\rho)}{2}) \nonumber\\
       &=& \rho + \frac{m}{2} + (\rho - \frac{m}{2})\, \gamma_5\, \varepsilon (\gamma_5 D_w(\rho)).
\end{eqnarray}
To accommodate the $SU(3)$ chiral transformation, it is usually convenient to use the chirally regulated field
$\hat{\psi} = (1 - \frac{1}{2} D_{ov}) \psi$ in lieu of $\psi$ in the interpolation field and operators.
That is to say, our valence quark propagator is
\begin{equation}
G \equiv D_{eff}^{-1} \equiv (1 - \frac{D_{ov}}{2}) D^{-1}_m = \frac{1}{D_c + m},
\end{equation}
where $D_c = \frac{\rho D_{ov}}{1 - D_{ov}/2}$ is chiral, i.e. $\{\gamma_5, D_c\}=0$~\cite{Chiu:1998gp}.

The overlap valence quark masses in lattice units are given in Tab.~\ref{tab:24_32}.
The corresponding pion masses range from 220 to 600 MeV.
\begin{table}
\begin{center}
\caption{The overlap valence quark masses $am_q$ in lattice units used in this analysis.}
\begin{tabular}{cccccccc}
\hline\hline
0.00620 & 0.00809 & 0.01020 & 0.01350 & 0.01720 & 0.02430 & 0.03650 & 0.04890 \\
\hline\hline
\end{tabular}
\label{tab:24_32}
\end{center}
\end{table}
Our quark propagators are calculated by using a point source on each configuration.
The numbers of configurations used in this work are given in Tab.~\ref{tab:confs}.
For three of the valence quark masses ($0.01350$, $0.02430$, $0.04890$) on ensemble c005, eight point sources on each configuration are used.
For the same three quark masses on ensemble c02, eight point sources are used on half of the 138 configurations.
The eight point sources are evenly distributed on the time slides and randomly distributed in 3-space from configuration to configuration
to reduce autocorrelations. Part of these propagators were calculated and used 
in the computation of renormalization constants~\cite{Liu:2013yxz} and
in the study of diquarks~\cite{Bi:2015ifa}.
We average the quark propagators from the eight sources on each configuration for these three valence quark masses.
Then together with the data from other configurations for other quark masses
a Jackknife procedure (one configuration eliminated each time) is done to get the statistical uncertainties in our analysis below.
Since ensemble c01 has the least statistics, the result from it will have the largest uncertainty.
While c005 will have the smallest statistical uncertainty.

Anti-periodic and periodic boundary conditions are used respectively in the time and spacial directions. Therefore the momentum modes are
\begin{equation}
ap=(\frac{2\pi k_1}{L}, \frac{2\pi k_2}{L}, \frac{2\pi k_3}{L}, \frac{(2k_4+1)\pi}{T}),
\end{equation}
where $k_\mu=-6,-5,...,6$.
To reduce Lorentz noninvariant discretization effects,
we use the momentum modes close to the diagonal line. This is achieved by doing a cut as was done in Ref.~\cite{Liu:2013yxz}
\begin{equation}
\frac{p^{[4]}}{(p^2)^2}<0.32,\quad\mbox{where } p^{[n]}=\sum_{\mu=1}^4 p_\mu^{n},\quad p^2=\sum_\mu p_\mu^2.
\label{eq:p4p22}
\end{equation}

\section{Analysis and discussions}
\label{sec:analysis}
\begin{figure}
\begin{center}
\includegraphics[height=2in,width=0.49\textwidth]{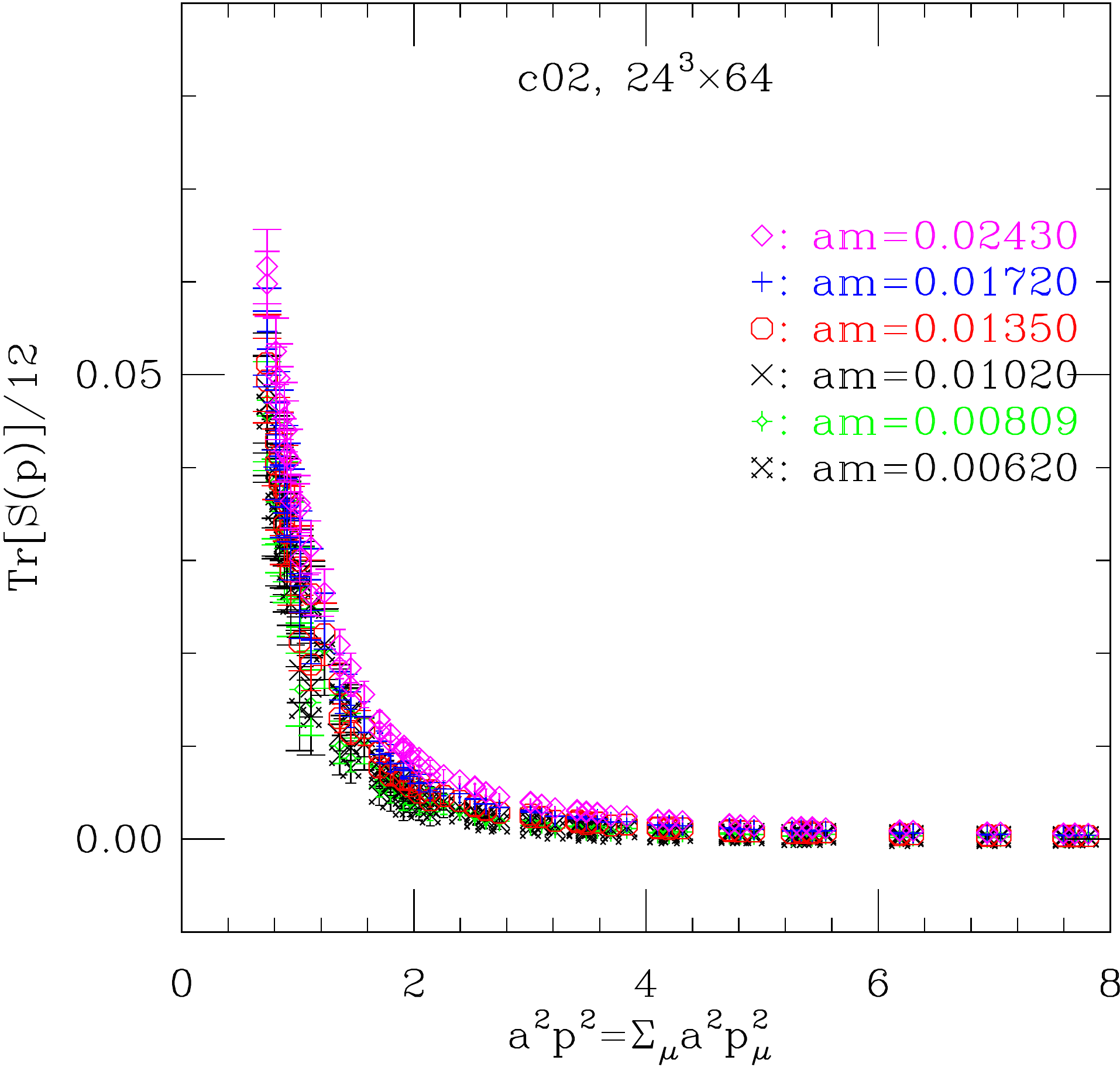}
\includegraphics[height=2in,width=0.49\textwidth]{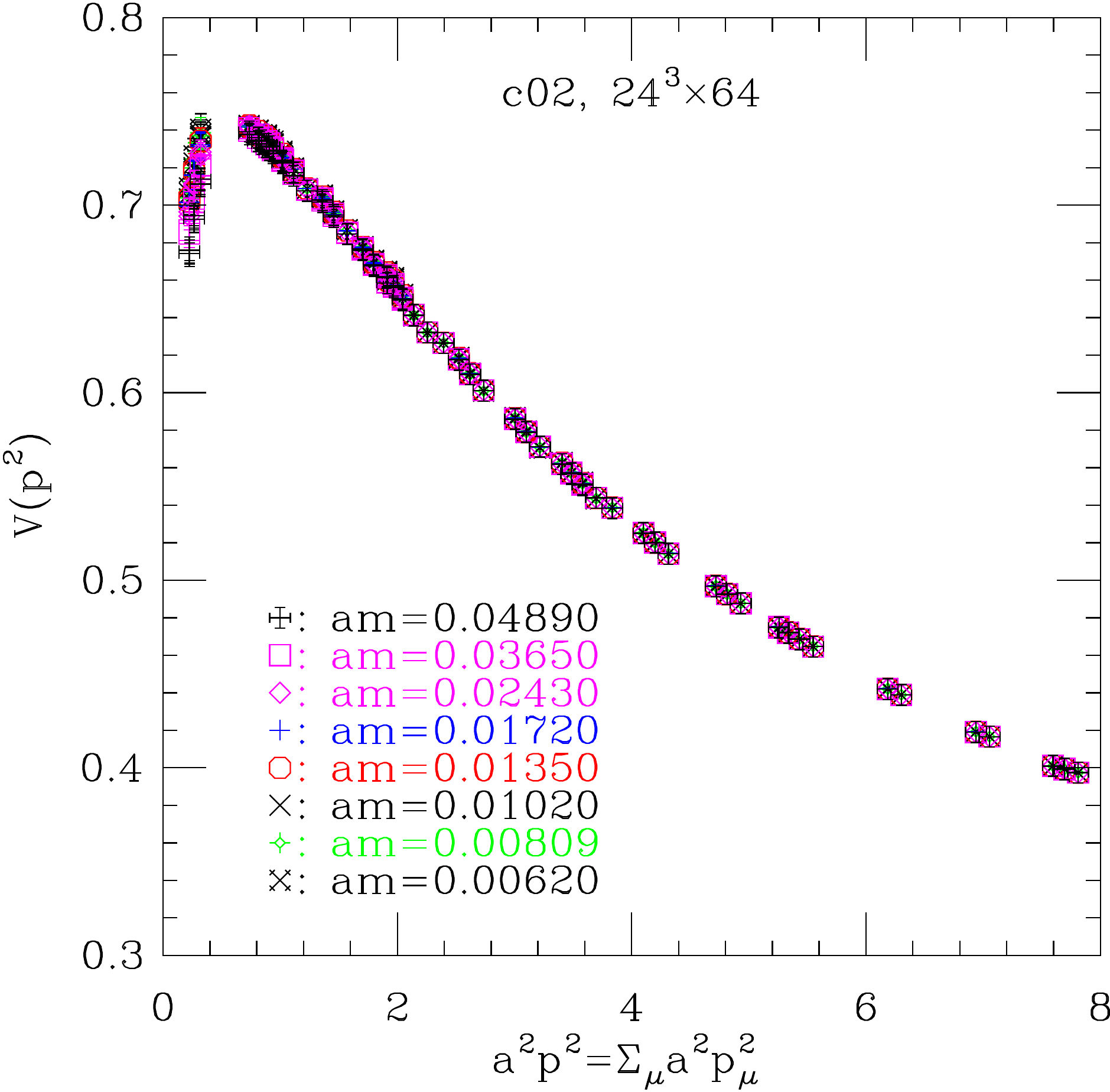}
\end{center}
\caption{Left: The bare vector form factor $S(p^2)$ divided by $a^2p^2$ from the quark propagators for various valence quark masses.
Right: The bare vector form factor $V(p^2)$ for various valence quark masses.}
\label{fig:ff_bare}
\end{figure}

From Eq.(\ref{eq:qprop}) we have
\begin{equation}
\frac{1}{12}\Tr[S_q(p)]=\frac{S(p^2)}{p^2},\quad\quad
\frac{1}{12}\Tr[i\slashed p S_q(p)]=V(p^2).
\end{equation}
In Fig.~\ref{fig:ff_bare} we show the bare scalar and vector form factors ($S(p^2)/p^2$ and $V(p^2)$) in lattice units
from our data ensemble c02 as functions of $a^2p^2$.
The scalar form factor has a visible quark mass dependence as is shown
in the graph on the left. On the contrary, the vector form factor in the graph on the right
has no visible quark mass dependence even at quite low $p^2$ region.
For example, at $a^2p^2=1.114$ the vector form factors for $am_q=0.00620$ and $0.04890$ agree with each other within the statistical uncertainties
(0.720(4) versus 0.717(3)).
This indicates the contribution from the $m_q^2$ term is quite small in Eq.(\ref{eq:VR}).
Therefore we can also expect the $m_q^3$ term in Eq.(\ref{eq:SR}) is negligible. Indeed at below we will see the quark mass dependence of
the scalar form factor can be well described by a linear function.

In our analysis below we take into account the reduced $\mathcal{O}(a^2p^2)$ discretization effects
by adding a term proportional to $a^2p^2$ in the fitting functions. However there are other artifacts of $\mathcal{O}(a^2g^2)$.
In Ref.~\cite{Burger:2012ti} the authors find that $\mathcal{O}(a^2g^2)$ effects are substantial in the vector form factor $V$, but modest in the ratio
$S/V$. Since we have not computed the lattice artifacts of $\mathcal{O}(a^2g^2)$ and thus can not remove them from our form factors, we
estimate their effects in our results by comparing the chiral condensates obtained
from analyzing the ratio of the form factors and from analyzing the scalar form factor alone.
The difference in the results from the two analysis will be taken as a systematic uncertainty.

\subsection{Analysis of the ratio of scalar to vector form factor}
\label{sec:ratio}
Since the ratio is expected to have much smaller $\mathcal{O}(a^2g^2)$ lattice artifacts than the scalar form factor,
we trust more on the chiral condensate from the analysis of the ratio.
The number from this analysis will be taken as our final result. 

The gluon condensate $\langle A^2\rangle$ in Landau gauge was determined in,
for example, Refs.~\cite{Blossier:2010vt,Blossier:2010ky}. In the analysis of Ref.~\cite{Burger:2012ti},
a compatible value of $\langle A^2\rangle$ was found but it seemed not yet stable against the order in perturbation theory.
Since $\langle A^2\rangle/p^2<\sim0.6\mbox{ GeV}^2/4\mbox{ GeV}^2=0.15$ is small in the range of $p^2$ in our following analysis
and the corresponding Wilson coefficient $|C_{A^2}|$ is also small ($\sim0.3$), 
we ignore the contribution from this condensate in Eq.(\ref{eq:VR}) as a first step
(note $V_{PT}$ is of order 1).
To obtain information about $\langle A^2\rangle$ from analyzing the vector form factor, 
we need more statistics and need to subtract the $\mathcal{O}(a^2g^2)$ artifacts.

The quark mass dependence of the vector form factor is quite small as was seen in Fig.~\ref{fig:ff_bare}.
This indicates we can keep only the first term
on the right hand side of Eq.(\ref{eq:VR}) in analyzing our data. Thus from Eqs.(\ref{eq:SR},\ref{eq:VR}), we have
for the bare and renormalized form factors
\begin{equation}
\frac{S_0}{V_0}=\frac{S_R}{V_R}=\frac{S_{PT}}{V_{PT}}m(\mu)+\frac{C_{\bar\psi\psi}(\mu,p^2)}{p^2V_{PT}}\langle\bar\psi\psi\rangle.
\end{equation}
Here the quark field renormalization constants $Z_q$ in the numerator and denominator cancel each other.

Define a ratio
\begin{equation}
R\equiv\frac{\Tr[S_q(p)]}{\Tr[i\slashed p S_q(p)]}=\frac{S(p^2)}{p^2V(p^2)},
\end{equation}
then in the chiral limit we have
\begin{equation}
\lim_{m_q\rightarrow0}R=\frac{C_{\bar\psi\psi}(\mu,p^2)}{(p^2)^2V_{PT}}\langle\bar\psi\psi\rangle.
\end{equation}
In lattice units and taking into account $\mathcal{O}(a^2p^2)$ lattice artifacts in the quark propagator, we use the following function
\begin{equation}
\lim_{m_q\rightarrow0}\frac{R}{a}=\frac{C_{\bar\psi\psi}(\mu,p^2)}{(a^2p^2)^2 V_{PT}}a^3\langle\bar\psi\psi\rangle+Ba^2p^2.
\label{eq:R_massless}
\end{equation}
to fit the ratio obtained from our lattice quark propagator. The dimensionless quark chiral condensate $a^3\langle\bar\psi\psi\rangle$
and $B$ are two fit parameters.

\begin{figure}
\begin{center}
\includegraphics[height=2in,width=0.49\textwidth]{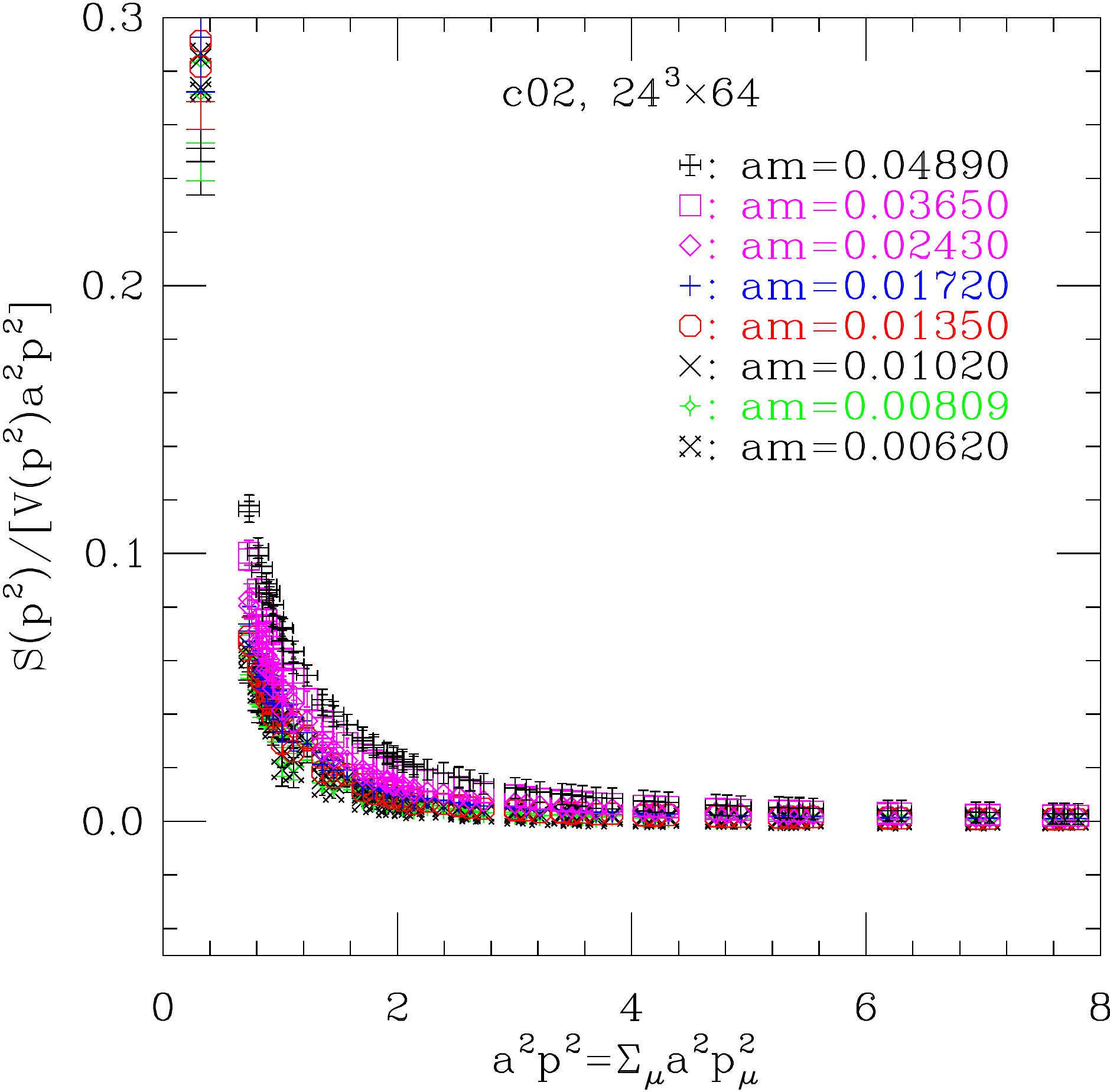}
\includegraphics[height=2in,width=0.49\textwidth]{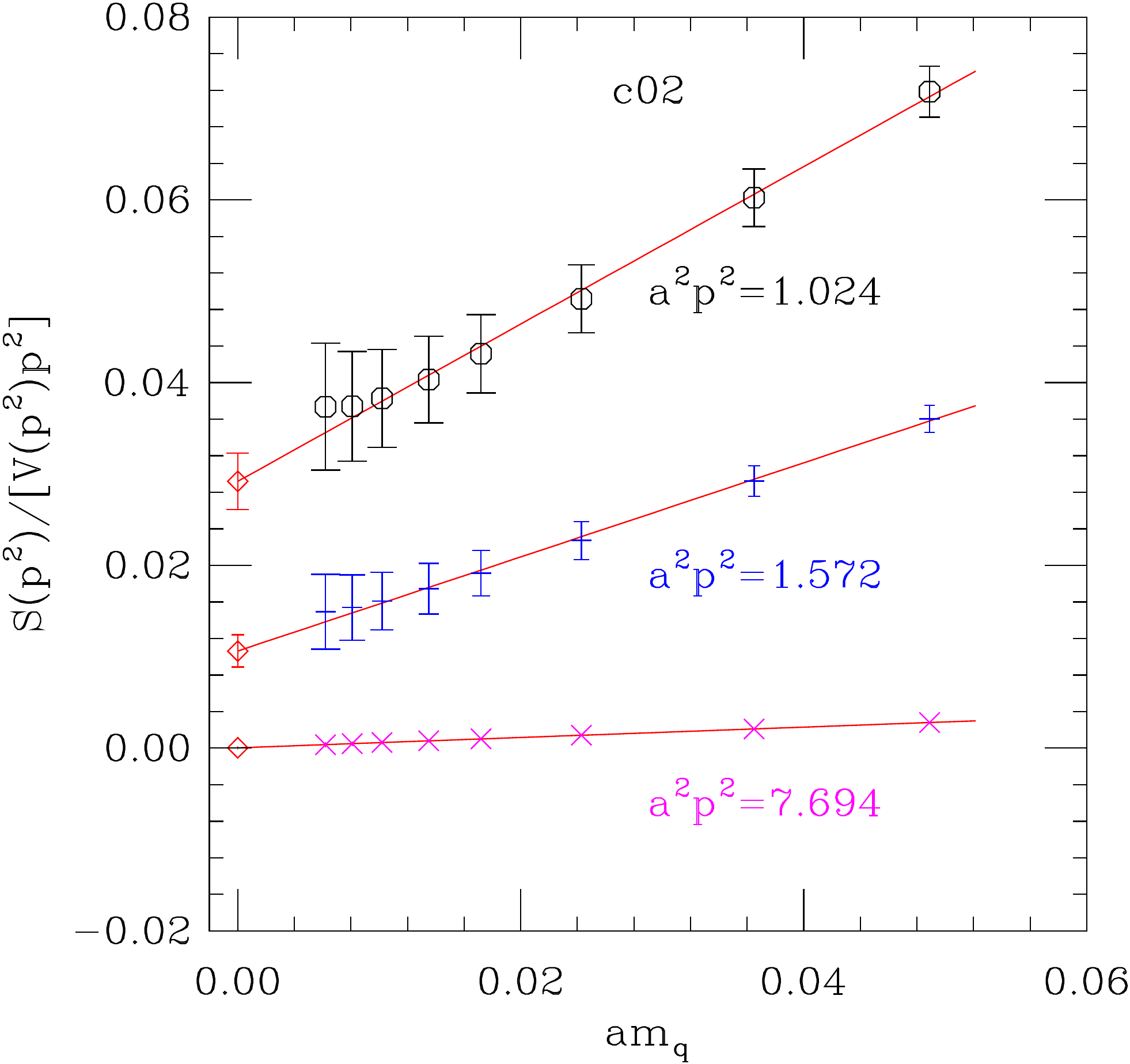}
\end{center}
\caption{Left: The ratio of form factors for various valence quark masses.
Right: Examples of linear extrapolation to the quark massless limit at three typical momentum values.}
\label{fig:R}
\end{figure}

In the graph on the left of Fig.~\ref{fig:R} we show the ratio $R$ (in lattice units)
as a function of $a^2p^2$ from ensemble c02 for various valence quark masses.
The graph on the right of Fig.~\ref{fig:R} shows examples of the linear chiral extrapolation of $R$ at three typical
momentum values: $a^2p^2=1.024$, $1.572$ and $7.694$. At all momentum values in our data for $R$ we see a good linear dependence on $am_q$.

\begin{figure}
\begin{center}
\includegraphics[height=2in,width=0.49\textwidth]{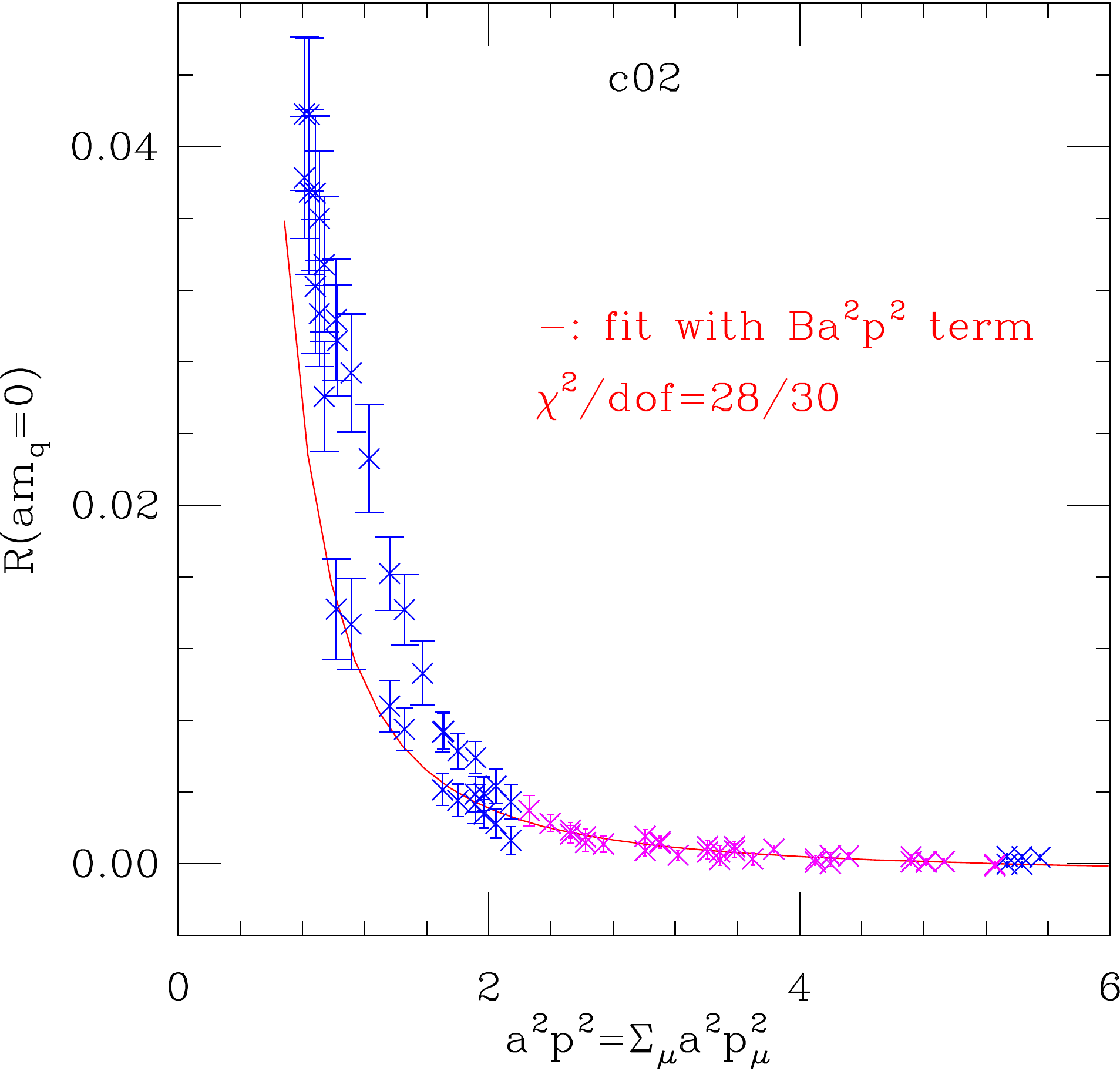}
\includegraphics[height=2in,width=0.49\textwidth]{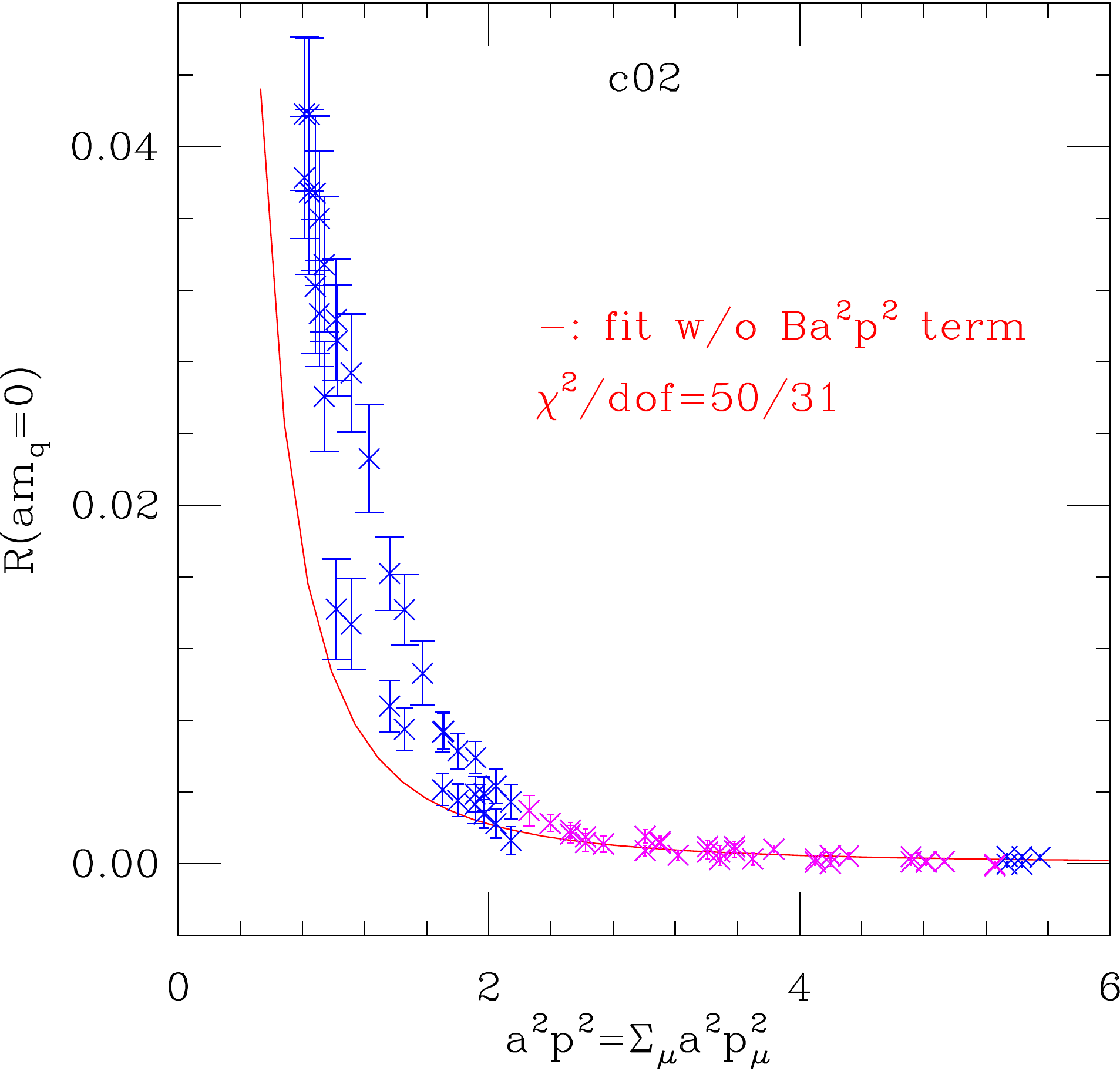}
\end{center}
\caption{Left: Fitting of the ratio $R$ in the valence quark massless limit.
Right: The same fit as in the left graph but without the $Ba^2p^2$ term. The data points in purple ($a^2p^2\in[2.2,5.3]$) are included in the fittings.}
\label{fig:fit_R_a2p2}
\end{figure}

Then we fit the ratio $R$ in the chiral limit to the function Eq.(\ref{eq:R_massless}).
Fig.~\ref{fig:fit_R_a2p2} shows an example of the fitting using a fitting range $a^2p^2\in[2.2,5.3]$. The fitting in the right graph
does not include the $Ba^2p^2$ term. Comparing it with the fitting in the left graph which does contain this term, we see that
the $Ba^2p^2$ term decreases $\chi^2$/dof significantly.
In evaluating the Wilson coefficients in the fitting function,
we use $\Lambda_{QCD}^\msbar=332(17)$ MeV for three flavors in the $\msbar$ scheme~\cite{Olive:2016xmw} to
compute the strong coupling constant $\alpha_s$. $\alpha_s^\msbar(2\mbox{ GeV})$ is calculated by using
its perturbative running to 3-loops since the Wilson coefficients are only known to 3-loops.
From this fitting, we get $\langle\bar\psi\psi\rangle^\msbar(2\mbox{ GeV})=-(307(37)(7)$ MeV$)^3$ by using the lattice spacing $1/a=1.75(4)$ GeV~\cite{Yang:2014sea}.
Here the first uncertainty is statistical and the second is from the uncertainty in the lattice spacing.

To check the stability of the result against the fitting range, we vary the lower and upper limits of $a^2p^2$.
In Tab.~\ref{tab:fit_range}, we give the $\chi^2$/dof of the fittings and the results of $\langle\bar\psi\psi\rangle$ against these changes.
As we see from the table, we can get a stable value for $\langle\bar\psi\psi\rangle$.
\begin{table}
\begin{center}
\caption{$\langle\bar\psi\psi\rangle^\msbar(2$ GeV) from fittings of $R$ with different fitting ranges on ensemble c02.
The first uncertainty is statistical and the second is from the uncertainty in the lattice spacing.}
\begin{tabular}{cccl}
\hline\hline
$a^2p^2\in$ & $p^2\in$/GeV$^2$ & $\chi^2/$dof & $(\langle\bar\psi\psi\rangle)^{1/3}$/MeV  \\
\hline
$[2.2, 5.5]$ & $[6.7, 16.8]$ & $0.91$ & $-299(36)(7)$  \\
$[2.2, 5.3]$ & $[6.7, 16.2]$ & $0.71$ & $-307(37)(7)$  \\
$[2.2, 5.1]$ & $[6.7, 15.6]$ & $0.73$ & $-304(37)(7)$  \\
$[2.2, 4.9]$ & $[6.7, 15.0]$ & $0.76$ & $-305(38)(7)$  \\
$[2.2, 4.7]$ & $[6.7, 14.4]$ & $0.79$ & $-309(39)(7)$  \\
\hline
$[2.6, 5.3]$ & $[8.0, 16.2]$ & $0.65$ & $-299(41)(7)$  \\
$[2.4, 5.3]$ & $[7.4, 16.2]$ & $0.61$ & $-300(39)(7)$  \\
$[2.0, 5.3]$ & $[6.1, 16.2]$ & $0.77$ & $-310(35)(7)$  \\
$[1.8, 5.3]$ & $[5.5, 16.2]$ & $1.01$ & $-318(32)(7)$  \\
\hline\hline
\end{tabular}
\label{tab:fit_range}
\end{center}
\end{table}

We then check the truncation error from the perturbative expansion of the Wilson coefficients.
We repeat the fittings with Wilson coefficients and $\alpha_s$ being evaluated at 2-loops and 1-loop.
The resulted numbers from data ensemble c02 are collected in Tab.~\ref{tab:n123}.
\begin{table}
\begin{center}
\caption{$\langle\bar\psi\psi\rangle^\msbar(2$ GeV) from fittings of $R$ on ensemble c02 with different truncation order
($n$-loops) in evaluating the Wilson coefficients and $\alpha_s$. The fitting range is $a^2p^2\in [2.2, 5.3]$.}
\begin{tabular}{ccl}
\hline\hline
$n$ &  $\chi^2/$dof & $(\langle\bar\psi\psi\rangle)^{1/3}$/MeV  \\
\hline
1 &  $0.73$ & $-330(39)(8)$  \\
2 &  0.72 & $-315(38)(7)$ \\
3 & $0.71$ & $-307(37)(7)$  \\
\hline\hline
\end{tabular}
\label{tab:n123}
\end{center}
\end{table}
Taking the difference between the center values with $n=2$ and $n=3$ as a systematic error, we finally get
$\langle\bar\psi\psi\rangle^\msbar(2\mbox{ GeV})=(-307(37)(7)(8)\mbox{ MeV})^3$ on ensemble c02. This is collected in Tab.~\ref{tab:ratio_all}.

Similarly in Tab.~\ref{tab:c01} and Tab.~\ref{tab:c005} we give the results from various fitting ranges
on the other two ensembles c01 and c005 respectively.
The truncation effects in the Wilson coefficients and $\alpha_s$ are examined too.
The quark condensates from all three ensembles are listed in Tab.~\ref{tab:ratio_all}.
\begin{table}
\begin{center}
\caption{$\langle\bar\psi\psi\rangle^\msbar(2$ GeV) from various fitting ranges on ensemble c01.
The first uncertainty is statistical and the second is from the uncertainty in the lattice spacing.}
\begin{tabular}{cccl}
\hline\hline
$a^2p^2\in$ & $p^2\in$/GeV$^2$ & $\chi^2/$dof & $(\langle\bar\psi\psi\rangle)^{1/3}$/MeV  \\
\hline
$[1.8, 3.8]$ & $[5.5, 11.6]$ & $0.97$ & $-299(42)(7)$  \\
$[1.8, 3.6]$ & $[5.5, 11.0]$ & $0.99$ & $-300(43)(7)$  \\
$[1.8, 3.4]$ & $[5.5, 10.4]$ & $1.03$ & $-293(45)(7)$  \\
\hline
$[2.2, 3.8]$ & $[6.7, 11.6]$ & $0.87$ & $-304(53)(7)$  \\
$[2.0, 3.8]$ & $[6.1, 11.6]$ & $0.78$ & $-295(48)(7)$  \\
$[1.6, 3.8]$ & $[4.9, 11.6]$ & $1.31$ & $-322(32)(7)$  \\
\hline\hline
\end{tabular}
\label{tab:c01}
\end{center}
\end{table}

\begin{table}
\begin{center}
\caption{$\langle\bar\psi\psi\rangle^\msbar(2$ GeV) from various fitting ranges on ensemble c005.
The first uncertainty is statistical and the second is from the uncertainty in the lattice spacing.}
\begin{tabular}{cccl}
\hline\hline
$a^2p^2\in$ & $p^2\in$/GeV$^2$ & $\chi^2/$dof & $(\langle\bar\psi\psi\rangle)^{1/3}$/MeV  \\
\hline
$[1.4, 3.3]$ & $[4.3, 10.1]$ & $1.15$ & $-302(11)(7)$  \\
$[1.4, 3.1]$ & $[4.3, 9.5]$ & $0.90$ & $-306(11)(7)$  \\
$[1.4, 2.9]$ & $[4.3, 8.9]$ & $0.98$ & $-306(12)(7)$  \\
\hline
$[1.8, 3.1]$ & $[5.5, 9.5]$ & $0.81$ & $-303(16)(7)$  \\
$[1.6, 3.1]$ & $[4.9, 9.5]$ & $0.73$ & $-300(13)(7)$  \\
$[1.2, 3.1]$ & $[3.7, 9.5]$ & $0.94$ & $-310(10)(7)$  \\
\hline\hline
\end{tabular}
\label{tab:c005}
\end{center}
\end{table}

\begin{table}
\begin{center}
\caption{$\langle\bar\psi\psi\rangle^\msbar(2$ GeV) on the three ensembles. The first uncertainty is statistical.
The second one is from the uncertainty of the lattice spacing. The third one is an estimation of the truncation effects
in the evaluations of the Wilson coefficients and $\alpha_s$.}
\begin{tabular}{ccccl}
\hline\hline
ensemble & $a^2p^2\in$ & $p^2\in$/GeV$^2$ & $\chi^2/$dof & $(\langle\bar\psi\psi\rangle)^{1/3}$/MeV  \\
\hline
c02 & $[2.2, 5.3]$ & $[6.7, 16.2]$ & $0.71$ & $-307(37)(7)(8)$  \\
c01 & $[1.8, 3.8]$ & $[5.5, 11.6]$ & $0.97$ & $-299(42)(7)(11)$  \\
c005& $[1.4, 3.1]$ & $[4.3, 9.5]$ & $0.90$ & $-306(11)(7)(13)$  \\
\hline\hline
\end{tabular}
\label{tab:ratio_all}
\end{center}
\end{table}

We also tried to do fittings in a same momentum range $a^2p^2\in [2.0, 3.2]$ on all three ensembles. What we found are
given in Tab.~\ref{tab:same_range}.
\begin{table}
\begin{center}
\caption{$\langle\bar\psi\psi\rangle^\msbar(2$ GeV) from a same fitting window $a^2p^2\in [2.0, 3.2]$ on all three ensembles.
The three uncertainties are as explained in Tab.~\ref{tab:ratio_all}.}
\begin{tabular}{ccl}
\hline\hline
ensemble &  $\chi^2/$dof & $(\langle\bar\psi\psi\rangle)^{1/3}$/MeV  \\
\hline
c02 &  $0.78$ & $-330(39)(8)(12)$  \\
c01 &  $0.79$ & $-281(57)(6)(10)$ \\
c005 & $0.85$ & $-303(18)(7)(12)$  \\
\hline\hline
\end{tabular}
\label{tab:same_range}
\end{center}
\end{table}

Besides all the above, we repeat the fittings with $\langle A^2\rangle^\msbar(2\mbox{ GeV})$ 
being fixed to 0.6 GeV$^2$ in Eq.(\ref{eq:VR}).
The resulted changes in $(\langle\bar\psi\psi\rangle)^{1/3}$ are 3 to 4 MeV, much smaller than the statistical or other uncertainties.
This means it is safe to ignore the contribution from $\langle A^2\rangle$ with our current statistics.

\begin{figure}
\begin{center}
\includegraphics[height=1.8in,width=2.0in]{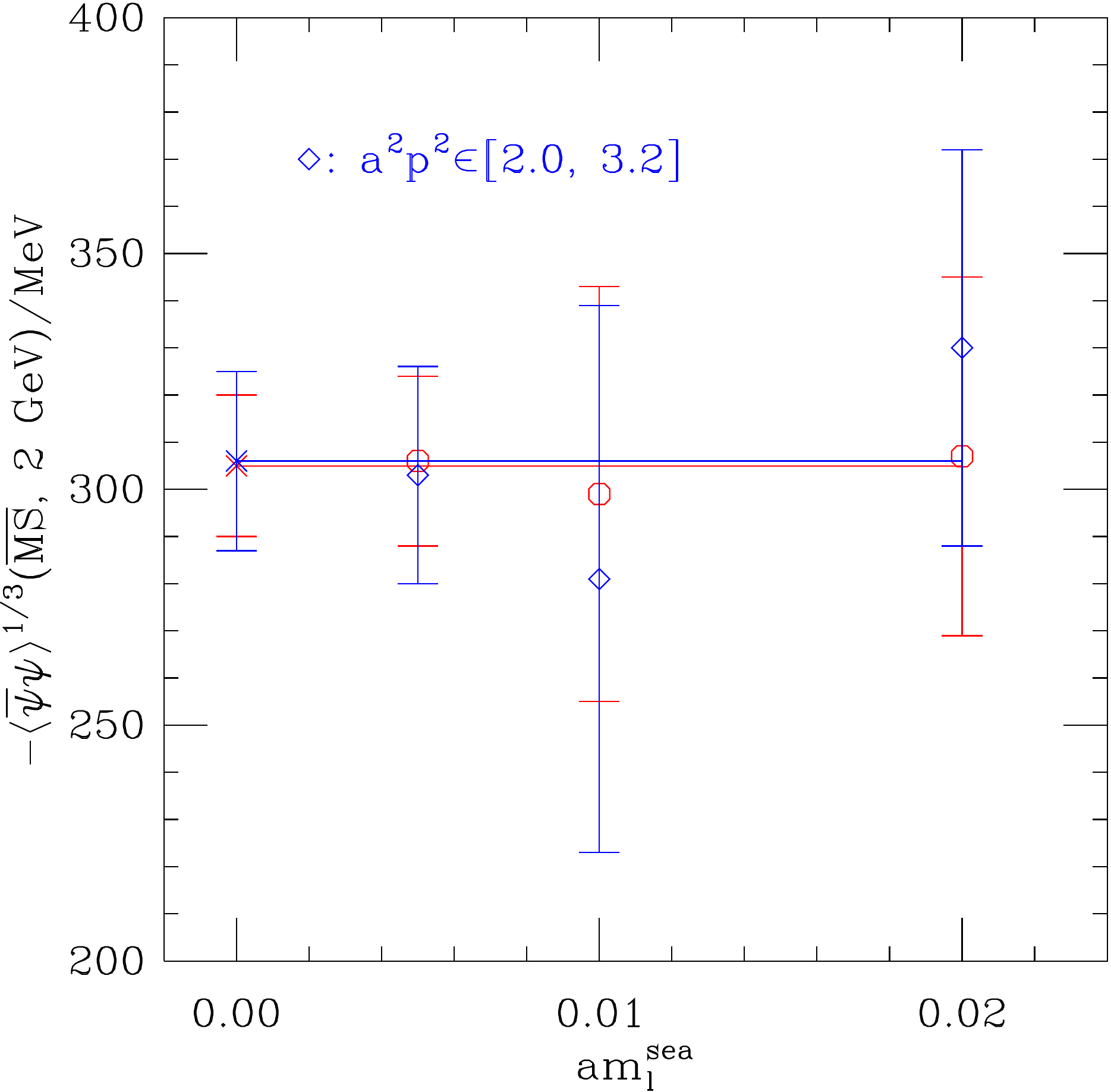}
\end{center}
\caption{The light sea quark mass dependence of $\langle\bar\psi\psi\rangle^\msbar(2$ GeV). The crosses are from constant fits.}
\label{fig:sea_mass}
\end{figure}

The light sea quark mass dependence is shown in Fig.~\ref{fig:sea_mass}, where we plot together the results from all three ensembles.
The three red points are those in Tab.~\ref{tab:ratio_all} from fittings with different $p^2$ range on each ensemble. 
The blue ones are those in Tab.~\ref{tab:same_range} from fittings
in a same momentum range on all three ensembles. In this graph, we have quadratically combined together the three uncertainties 
in Tab.~\ref{tab:ratio_all} and Tab.~\ref{tab:same_range} respectively.
Since we do not see an apparent sea quark mass dependence with our relatively large uncertainties,
we do a constant fit to finally obtain
\begin{equation}
\langle\bar\psi\psi\rangle^\msbar(2\mbox{ GeV})=(-305(15)\mbox{ MeV})^3  \quad(\text{different fitting ranges}),
\label{eq:cond_ratio}
\end{equation}
and
\begin{equation}
\langle\bar\psi\psi\rangle^\msbar(2\mbox{ GeV})=(-306(19)\mbox{ MeV})^3  \quad(\text{same fitting range}).
\end{equation}
These two numbers are in good agreement with each other.

\subsection{Analysis of the Scalar form factor}
There may be non-negligible $\mathcal{O}(a^2g^2)$ lattice artifact in our scalar 
and vector form factors as were seen in Ref.~\cite{Burger:2012ti}
with Wilson twisted mass fermions. At large $p^2$, difference was seen 
in the $\mathcal{O}(a^2g^2)$-corrected and un-corrected vector
form factor~\cite{Burger:2012ti}. Unfortunately, We have not calculated this artifact yet and therefore could not do this correction to our data. 
To estimate its effects, we analyze the scalar form factor in the chiral limit to
obtain the chiral condensate and compare the result with the one from Sec.~\ref{sec:ratio}.

From Eq.(\ref{eq:SR}) we see in the chiral limit the scalar form factor is related to the chiral condensate by
\begin{equation}
S_R(p^2) = \frac{C_{\bar\psi\psi}(\mu,p^2)}{p^2}\langle\bar\psi\psi\rangle(\mu).
\label{eq:SR_massless}
\end{equation}
With a quark field renormalization
constant $\psi_R=Z_q^{1/2}\psi$ and taking into account $\mathcal{O}(a^2p^2)$ effects, we have
\begin{equation}
\frac{1}{12}\Tr[S_q(a,p)]=\frac{S_0(a^2p^2)}{a^2p^2}=\frac{S_R(a^2p^2)}{Z_q a^2p^2}
= \frac{C_{\bar\psi\psi}(\mu,p^2)}{Z_q (a^2p^2)^2}a^3\langle\bar\psi\psi\rangle(\mu)+Ba^2p^2.
\label{eq:S_massless}
\end{equation}
Here we have put everything in lattice units. 
Thus the quark field renormalization constant $Z_q$ is needed in the analysis of the scalar form factor. 

\subsubsection{Quark field renormalization}
\label{sec:zq}
\begin{figure}
\begin{center}
\includegraphics[height=1.8in,width=2.0in]{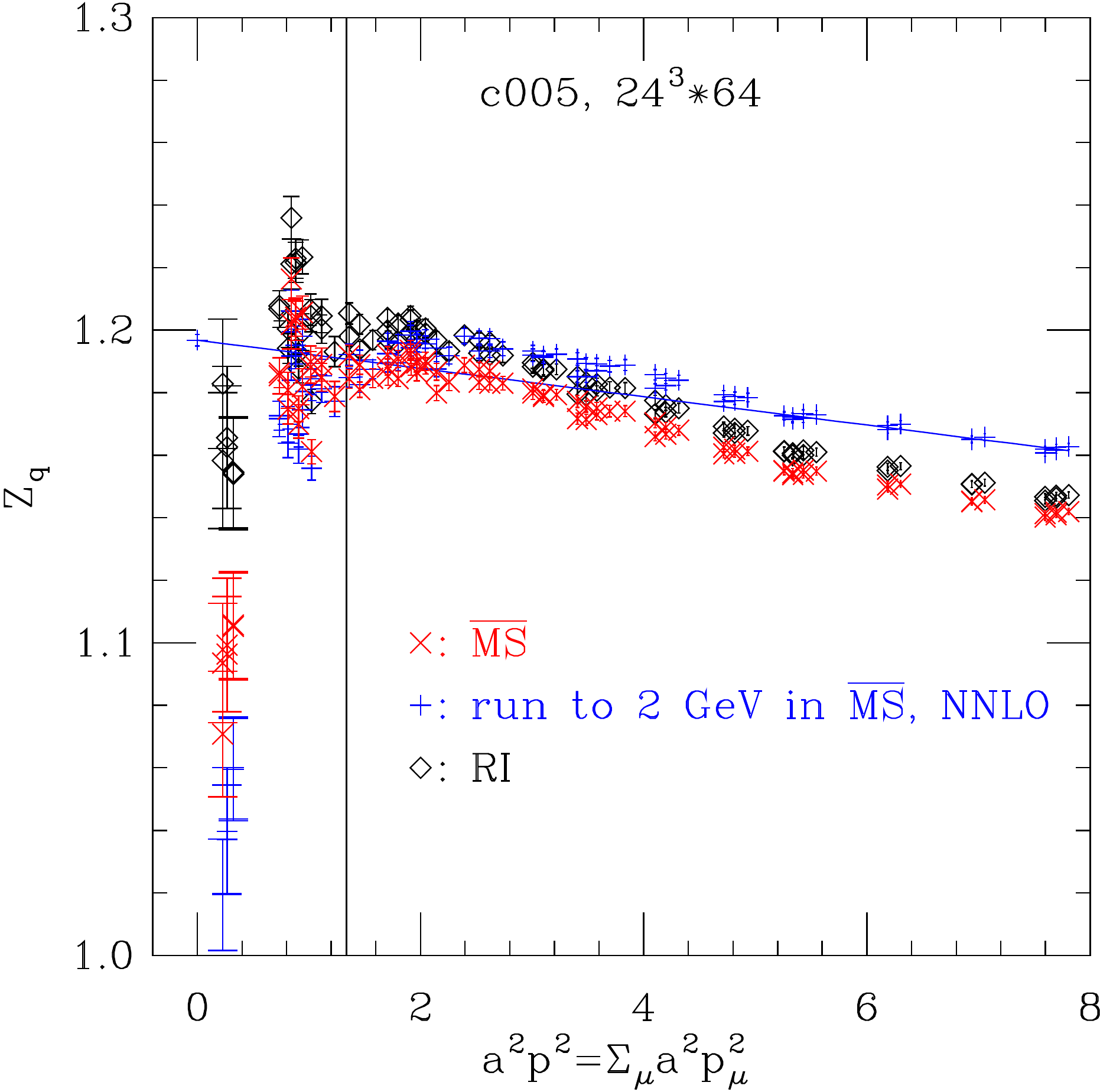}
\end{center}
\caption{The quark field renormalization constant for ensemble c005. The black vertical line indicates the position of $\mu=2$ GeV.}
\label{fig:zq}
\end{figure}

Our $Z_q$ is first calculated in the
RI-MOM scheme~\cite{Martinelli:1994ty} and then converted to the $\msbar$ scheme.
The detailed calculation in the RI-MOM scheme for our work can be found in Ref.~\cite{Liu:2013yxz}.
We first use the axial vector Ward Identity to obtain $Z_A^{WI}$, which equals to $Z_A$ in the RI-MOM scheme.
Then from it $Z_q$ in the RI-MOM scheme is computed at several valence quark masses.
The results of $Z_q$ show little quark mass dependence (see Fig.3 in Ref.~\cite{Liu:2013yxz}).
We now do a linear extrapolation of $Z_q$ in the quark mass to the chiral limit. 
The results in this limit are shown by the black diamonds in Fig.~\ref{fig:zq}.
Then the conversion ratio calculated by perturbation theory~\cite{Chetyrkin:1999pq}
to 3-loops is used to get $Z_q$ in the $\msbar$ scheme, which is shown by the red crosses in Fig.~\ref{fig:zq}.
After running $Z_q^\msbar$ from
an initial scale $p^2$ to $\mu^2=(2$ GeV$)^2$ by using its anomalous dimension to 3-loops, we obtain the blue pluses in Fig.~\ref{fig:zq}.
The deviation of the blue pluses from a constant at large initial scales is attributed to
$\mathcal{O}(a^2p^2)$ lattice artifacts. Thus a linear extrapolation in $a^2p^2$ to $a^2p^2=0$ is done to get $Z_q^\msbar(2\mbox{ GeV})$
(illustrated by the blue line using data points at $a^2p^2>5$).

The results of $Z_q$ in the $\msbar$ scheme are given in Tab.~\ref{tab:zq2GeV} for the three ensembles.
\begin{table}
\begin{center}
\caption{$Z_q^\msbar(2$ GeV) on the three ensembles. The first error is statistical and the second one
is a 1\% systematic error.}
\begin{tabular}{cccc}
\hline\hline
ensemble &   c02 & c01 & c005 \\ 
$Z_q^\msbar(2$ GeV) &  1.202(2)(12) & 1.209(3)(12) & 1.197(2)(12) \\ 
\hline\hline
\end{tabular}
\label{tab:zq2GeV}
\end{center}
\end{table}
Similarly to what have been done to $Z_S$ for the scalar density in Ref.~\cite{Liu:2013yxz} (see its Tab.~V),
we find a 1\% systematic uncertainty for $Z_q$ from
the uncertainty in the lattice spacing, the uncertainty in $\Lambda_{QCD}^\msbar$,
the truncation error of the perturbative conversion ratio and the variation of the fitting range in
the $a^2p^2$ linear extrapolation. This systematic uncertainty is given in Tab.~\ref{tab:zq2GeV}.

\subsubsection{Fitting results}
\begin{figure}
\begin{center}
\includegraphics[height=2in,width=0.49\textwidth]{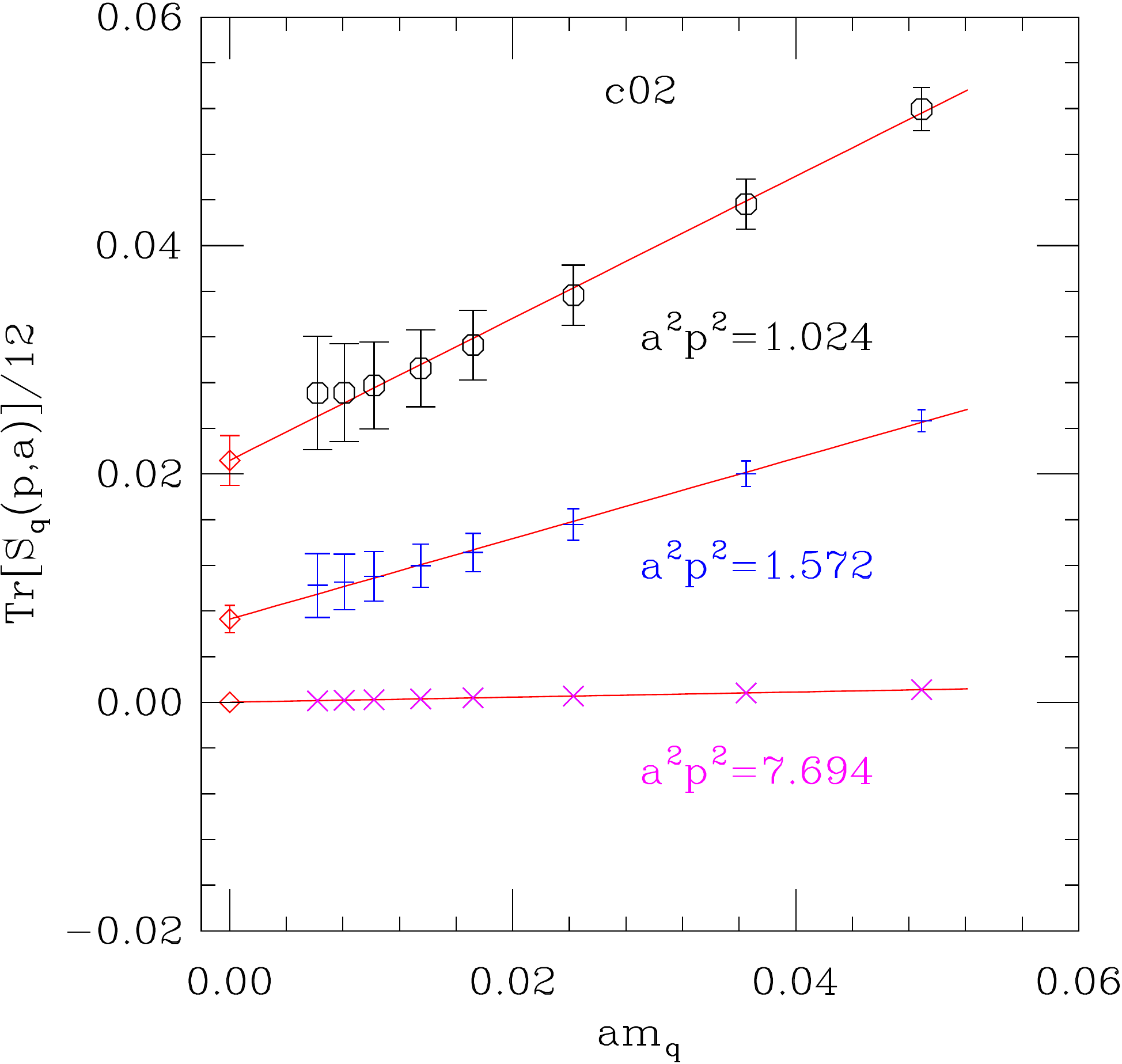}
\includegraphics[height=2in,width=0.49\textwidth]{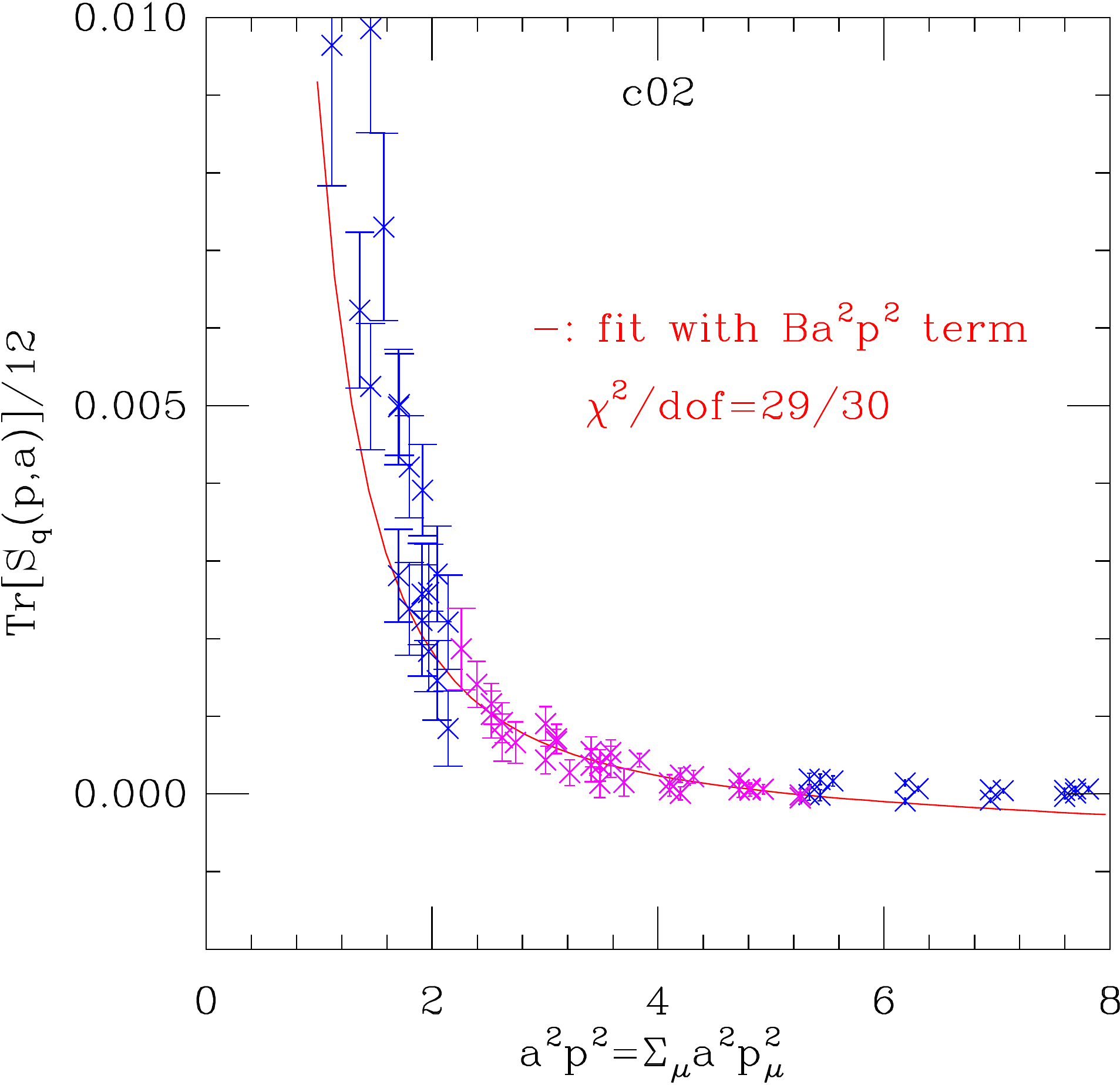}
\end{center}
\caption{Left: Examples of linear extrapolation of the scalar form factor to the quark massless limit at three typical momentum values.
Right: Fitting of the scalar form factor in the chiral limit to Eq.(~\ref{eq:S_massless}) on ensemble c02.}
\label{fig:scalar}
\end{figure}

We have shown the scalar form factor (divided by $p^2$) in the left graph of Fig.~\ref{fig:ff_bare}.
In Fig.~\ref{fig:scalar} we show the linear chiral extrapolation of the scalar form factor and the fitting of the
chiral limit results to
the function Eq.(\ref{eq:S_massless}).
Again, we find in the fit the $Ba^2p^2$ term decreases $\chi^2/$dof significantly. As was done in the analysis of the ratio of form factors
in Sec.~\ref{sec:ratio}, we check the stability of the results of $\langle\bar\psi\psi\rangle$ against the fitting range in $p^2$,
and against the order of
truncation in the evaluations of the Wilson coefficients and $\alpha_s$.
Since the uncertainty of $Z_q$ is quite small compared with other sources of uncertainties,
we have ignored its propagation to the uncertainty of the quark chiral condensate.

The results of $\langle\bar\psi\psi\rangle$ from the three ensembles are given in Tabs.~\ref{tab:scalar_all},\ref{tab:scalar_same_range}. 
Tab~\ref{tab:scalar_all} is for fittings
with different $p^2$ window on each ensemble and Tab~\ref{tab:scalar_same_range} for fittings with a same $p^2$ window on all three ensembles.
They are in agreement within errors.
\begin{table}
\begin{center}
\caption{$\langle\bar\psi\psi\rangle^\msbar(2$ GeV) on the three ensembles from the analysis of the scalar form factor. 
The first uncertainty is statistical.
The second one is from the uncertainty of the lattice spacing. The third one is an estimation of the truncation effects
in the evaluations of the Wilson coefficients and $\alpha_s$.}
\begin{tabular}{ccccl}
\hline\hline
ensemble & $a^2p^2\in$ & $p^2\in$/GeV$^2$ & $\chi^2/$dof & $(\langle\bar\psi\psi\rangle)^{1/3}$/MeV  \\
\hline
c02 & $[2.2, 5.3]$ & $[6.7, 16.2]$ & $0.76$ & $-272(32)(6)(7)$  \\
c01 & $[1.8, 3.8]$ & $[5.5, 11.6]$ & $1.02$ & $-278(35)(6)(9)$  \\
c005& $[1.4, 3.1]$ & $[4.3, 9.5]$ & $1.03$ & $-288(10)(7)(11)$  \\
\hline\hline
\end{tabular}
\label{tab:scalar_all}
\end{center}
\end{table}

\begin{table}
\begin{center}
\caption{$\langle\bar\psi\psi\rangle^\msbar(2$ GeV) from fittings of the scalar form factor in a same fitting window $a^2p^2\in [2.0, 3.2]$ 
on all three ensembles.
The three uncertainties are as explained in Tab.~\ref{tab:scalar_all}.}
\begin{tabular}{ccl}
\hline\hline
ensemble &  $\chi^2/$dof & $(\langle\bar\psi\psi\rangle)^{1/3}$/MeV  \\
\hline
c02 &  $1.00$ & $-299(35)(7)(9)$  \\
c01 &  $0.83$ & $-262(47)(6)(9)$ \\
c005 & $0.86$ & $-280(16)(6)(9)$  \\
\hline\hline
\end{tabular}
\label{tab:scalar_same_range}
\end{center}
\end{table}

The light sea quark mass dependence of $\langle\bar\psi\psi\rangle$ is again small compared with our uncertainties. 
Thus we do a constant fit similar to Fig.~\ref{fig:sea_mass} in Sec.~\ref{sec:ratio}. What we obtain are
\begin{equation}
\langle\bar\psi\psi\rangle^\msbar(2\mbox{ GeV})=(-284(13)\mbox{ MeV})^3 \quad  (\text{different fitting ranges}),
\label{eq:cond_scalar}
\end{equation}
and
\begin{equation}
\langle\bar\psi\psi\rangle^\msbar(2\mbox{ GeV})=(-282(16)\mbox{ MeV})^3 \quad  (\text{same fitting range}).
\end{equation}

\section{Summary}
\label{sec:summary}
We determine the quark chiral condensate by fitting lattice data of the overlap quark propagator to its operator product expansion
in the $\msbar$ scheme in Landau gauge.
We perform two analyses. One uses the ratio of scalar to vector form factor of the propagator, which is supposed to
have modest $\mathcal{O}(a^2g^2)$ lattice artifacts. The other one uses the scalar form factor. We use the result from the second analysis
to estimate the uncertainty from the $\mathcal{O}(a^2g^2)$ artifacts.
The fitting range of the momentum in our analysis is varied to check the stability of the results. The truncation error
in evaluating the Wilson coefficients and $\alpha_s$ is also examined. 
Three ensembles of 2+1-flavor domain wall fermion configurations are used to check the light sea quark
mass dependence.

We take the number in Eq.(\ref{eq:cond_ratio}) as our final result. The difference between the center values in Eq.(\ref{eq:cond_ratio})
and Eq.(\ref{eq:cond_scalar}) is taken as a systematic
uncertainty due to the $\mathcal{O}(a^2g^2)$ effects in our data. That is to say, our final result is
\begin{equation}
\langle\bar\psi\psi\rangle^\msbar(2\mbox{ GeV})=(-305(15)(21)\mbox{ MeV})^3.
\label{eq:final_result}
\end{equation}
Here the first error contains uncertainties from statistics, the lattice spacing and truncations in perturbative calculations
of the Wilson coefficients and $\alpha_s$.

Our result Eq.(\ref{eq:final_result}), with a relatively large error bar, agrees with the FLAG-3 average $\Sigma^{1/3}=274(3)$ MeV~\cite{Aoki:2016frl}
for $N_f=2+1$ flavor lattice calculations.
To improve our work, the $\mathcal{O}(a^2g^2)$ effects should be calculated and removed from the lattice data of the quark propagator.
With more statistics the effects of the $\langle A^2\rangle$ term can be checked carefully. 
Furthermore, calculations at more lattice spacings should be done to enable a continuum extrapolation.

\section*{Acknowledgements}
We thank the RBC-UKQCD Collaborations for sharing the domain wall fermion configurations. 
We also thank Andreas Maier and Konstantin Chetyrkin
for useful correspondence.
This work is partially supported by the National Science Foundation of China (NSFC) under Grants 11575196, 11575197 and 11335001. 
YC and ZL acknowledge the support of NSFC and DFG (CRC110). Part of the numerical computations are performed
on Tianhe-II at the National Supercomputer Center in Guangzhou.

\end{document}